\documentclass[]{aa}
\usepackage{natbib}
\usepackage{graphicx}
\usepackage{wasysym}
\usepackage{txfonts}
\def\EBV{\mbox{E$_{\rm B-V}$}}
\def\AF{\mbox{A$_{F\prime,F}$}}

\def\AV{\mbox{A$_{\rm V}$}}

\def\nH2{\mbox{${\rm n}_\HH}$}

\def\pccc{~{\rm cm}^{-3}} 
 
\def\pcc {~{\rm cm}^{-2}}

\def\Tsub#1 {\mbox{${\rm T}_{\rm #1}$}}
\def\TK  {\Tsub K }

\def\arcsec{\mbox{$^{\prime\prime}$}}

\def\degr{$^{\rm o}$}
\def\p{\mbox{$^+$}}

\def\THC{\mbox{{$^{13}$C}}} 
 
\def\chp{\mbox{CH\p}}
 
\def\cch{\mbox{C$_2$H}}
\def\cfh{\mbox{C$_4$H}}  
\def\hhco{\mbox{H$_2$CO}}
\def\h13cop{\mbox{{H$^{13}$CO\p}}}

\def\C3H{\mbox{C$_3$H}}
\def\c3h2{\mbox{C$_3$H$_2$}}
\def\cc3h2{\mbox{{\it c}-C$_3$H$_2$}}
 
 \def\R0{R$_0$}
\def\G0{\mbox{G$_0$}}

\def\ddeg{{}^\circ\kern-.1em}

\def\kms{\mbox{km\,s$^{-1}$}}
\def\ps{\mbox{s$^{-1}$}}
\def\bll{BL Lac}

\def\E#1 {$10^{#1}$}
\def\E#1 {E{#1}}
\def\P#1,{$\nH2\TK~=~#1\times~10^4\pccc$~K}
\def\ec#1,#2,#3,{#1\,(#2)\E{#3}}

\def\H3{\mbox{H$_3$}}

\def\RH2{\mbox{R$_{\rm G}$}}
\def\g13{\mbox{g$_{13}$}} 

\def\cc3h{\mbox{{\it c}-\C3H}}
\def\lc3h{\mbox{{\it l}-\C3H}}

\def\hthcop{\mbox{H$^{13}$CO\p}}

\newcommand{\emm}[1]{\ensuremath{#1}}   
\newcommand{\emr}[1]{\emm{\mathrm{#1}}} 

\newcommand{\hcop}{\emr{HCO^+}} 
 
\newcommand{\HH}{\emr{H_2}}

\renewcommand{\coth}{\emr{^{13}CO}}

\title{HCO, \cc3h\ and CF\p : three new molecules in diffuse, translucent and ``spiral-arm'' clouds}

\author{H. S. Liszt\inst{1}, J. Pety\inst{2,3}, M. Gerin\inst{4,3} \& R. Lucas\inst{5}}
\institute{National Radio Astronomy Observatory,
           520 Edgemont Road,
           Charlottesville, VA,
           USA 22903-2475
\and       Institut de Radioastronomie Millim\'etrique,
           300 Rue de la Piscine,
           F-38406 Saint Martin d'H\`eres,
           France
\and       Observatoire de Paris (CNRS UMR 8112), 
           61 av. de l'Observatoire, 75014, Paris, 
           France
\and       LERMA/LRA, Ecole Normale Sup\'erieure, 
           24 rue Lhomond, 75005 Paris, 
           France
\and       Institut Plan\'etologie et d'Astrophysique de Grenoble (UMR 5274),
           BP 53 F-38041 Grenoble Cedex 9, 
           France
}

\begin{document}
\date{received \today}%
\offprints{H. S. Liszt}%
\mail{hliszt@nrao.edu}%
%
\abstract
{}
{To observe molecular absorption from diffuse clouds across 3mm receiver band.}
{We used the EMIR receiver and FTS spectrometer at the IRAM 30m telescope to construct
absorption spectra toward bright extra-galactic background sources  at 195 kHz 
 spectral resolution ($\approx$ 0.6 \kms).  
 We used the IRAM Plateau de Bure interferometer to synthesize  
  absorption spectra of \hthcop\ and HCO toward the galactic HII region W49.}
{HCO, \cc3h\ and CF\p\ were detected toward 
 the blazars \bll\ and 3C111 having \EBV\ = 0.32 and 1.65 mag.  
  HCO was observed in absorption from ``spiral-arm'' clouds in the galactic plane 
  occulting W49.  The complement of detectable molecular species in the 
  85 - 110 GHz absorption spectrum of diffuse/translucent gas is now fully determined at rms 
  noise level $\delta_\tau \approx 0.002$ at \EBV\ = 0.32 mag (\AV\ = 1 mag) and 
 $\delta_\tau$/\EBV\ $\approx\ 0.003$ mag$^{-1}$ overall.
}
{As with OH, \hcop\ and \cch, the relative abundance of \cc3h\ varies little between
diffuse and dense molecular gas, with  N(\cc3h)/N({\it o-c}-\c3h2) $\approx$ 0.1.
We find N(CF\p)/N(H$^{13}$CO\p) $\approx 5$, N(CF\p)/N(\cch) $\approx$ 0.005-0.01 and
because N(CF\p) increases with \EBV\ and with the column densities of other
molecules we infer that fluorine remains in the gas phase as HF well beyond \AV\ = 1 mag.  
We find N(HCO)/N(H$^{13}$CO\p) = 16 toward \bll, 3C111 and the 40 km/s spiral arm cloud toward 
W49, implying X(HCO) $\approx 10^{-9}$, about 10 times higher than in dark clouds.
The behaviour of HCO is consistent with previous suggestions that it forms from 
C\p\ and \HH, even when \AV\ is well above 1 mag.  The survey can be used to place useful 
upper limits on some species, for instance N(\hhco)/N(\HH CS) $>$ 32 toward 3C111, compared
to 7 toward TMC-1, confirming the possibility of a gas phase formation route to \hhco.
In general, however, the hunt for new  species will probably be
more fruitful at cm- and sub-mm wavelengths for the near future.
}


\authorrunning{Liszt, Pety, Gerin \& Lucas} \titlerunning{3mm sweep}

\maketitle{}

%

\section{Introduction}

Microwave and sub-mm absorption-line spectroscopy have greatly extended the 
inventory of molecules known to exist in diffuse molecular interstellar 
clouds, that is, clouds with appreciable \HH\ content but \AV $\la 1$ mag.
Just in the past few years, sub-mm observations from the PRISMAS project 
using the HIFI instrument on HERSCHEL, from the APEX telescope in 
the Atacama  and the GREAT instrument on SOFIA have more than doubled 
the number of known species following
observation of the hydrides and hydride ions of oxygen, nitrogen, sulfur,
fluorine and chlorine \citep{GerLev+12,NeuFal+12}.  Sub-mm observations of the
familiar species CH \citep{GerdeL+10} and \chp\ \citep{FalGod+10,GodFal+12}
have allowed these species, previously seen only in the optical absorption 
spectra of relatively nearby stars, to be tracked across the disk of the 
Milky Way.

Even so, the search for new molecules and the overall effort to systematize the 
diffuse cloud chemistry have been seriously hindered by the narrow bandwidths 
that were available for high sensitivity observations of heavier polyatomic 
species in the  microwave (cm-wave and mm-wave) domain.  This impediment is 
gradually being
overcome by new technology such as the WIDAR correlator at the Karl Guthe Jansky 
VLA that we used to detect {\it l}-\c3h2\ and survey the abundance of several 
small hydrocarbons \citep{LisSon+12}.  An even larger development is the advent of
the very wide-band EMIR receivers at the IRAM 30m Telescope on Pico de Veleta, 
which produced the recent 1-3 mm WHISPER surveys of emission from the Horsehead nebula 
\citep{GuzRou+12,PetGra+12,GraPet+13}.

Here we describe a 3mm band survey of absorption from the galactic diffuse and translucent 
clouds seen toward the mm-bright blazars
\bll\ (\EBV\ = 0.32 mag) and 3C111 (\EBV\ = 1.65 mag).  As a result of this 
work the mm-wave absorption 
spectrum is now known at a detection level of 1\% absorption or
better over the 3mm band and this paper reports the detection of three new species
HCO, \cc3h\ and CF\p, toward both objects.

\begin{figure}
\includegraphics[height=15.5cm]{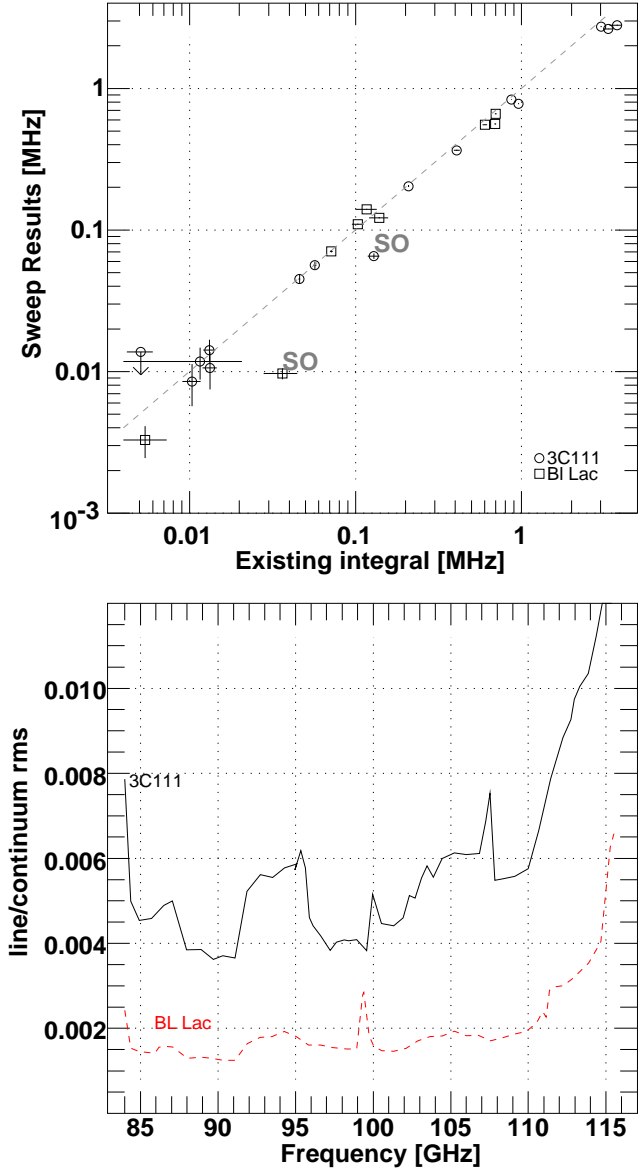}
  \caption[]{Properties of the spectral sweep.  Top: Line profile integrals 
  (equivalent widths) for the 
  spectral sweep plotted against those previously obtained by synthesis at the 
  Plateau de Bure Interferometer in our prior work. Bottom: 8-channel running mean
  channel-channel rms in the line/continuum ratio at 0-absorption for 3C111 and \bll.
  The points labelled SO in the upper panel represent the 3$_2$-2$_1$ line at 99.3 GHz 
  as discussed in Sect 2.1 of the text.  }
\end{figure}

The plan of this work is as follows.  The observations are described in Sect.
2.  In Sections 3-5 we discuss the detection and chemistry of HCO, \cc3h\ and CF\p, 
respectively.  Section 6 is the briefest of summaries.

\section{Observations and data reduction}

\subsection{The 3mm band EMIR spectral sweep at the 30m}

Each of three sources (see Table 1) was observed for some 
20 hours over the course of five days observing in 2012 August.  Only
the data from \bll\ (B2200+420)  and 3C111 (B0415+379) will be discussed here 
because the flux of B0355+508 (NRAO150) was too low to detect the new species 
that were seen toward the other sources.
We used the broadband EMIR receiver with an FTS spectrometer at 195 kHz resolution 
and channel separation (0.60 \kms\ at 100 GHz) while observing
simultaneously over both 8 GHz-wide sidebands, in much the same manner as 
described previously for the 1-3mm WHISPER surveys of emission from the edge-on PDR 
in the Horsehead nebula \citep{GuzPet+12}.
That is, we wobble-switched the secondary mirror symmetrically with an azimuth 
throw of 60\arcsec\ about the positions
of the blazar background sources and offset the receiver local oscillator
to produce continuous coverage, with some overlap, over very nearly the full 
receiver band.  In principle, observing the 3mm frequency band requires only
    two tunings. However, every frequency was observed with two different
 frequency settings separated by 500 MHz to allow us to remove
potential ghost images arising from  lines in the image side band,
given that the typical rejection of the EMIR sideband separating mixers in only 13 dB
(a factor 20).

The temperature scale was computed and applied in
the GILDAS/MIRA software on 4GHz chunks at a time. All other
  reduction steps were done in the GILDAS/CLASS software\footnote{See
 \texttt{http://www.iram.fr/IRAMFR/GILDAS} for more information about
 the GILDAS software~\citep{Pet05}.}. In brief, we fit the
 continuum  and divided the spectra by the continuum baseline to yield
 spectra as line/continuum ratios. We baselined the spectra with a low
 order baseline and rejected frequency ranges
 whose rms variation with time was much larger than the noise level,
  to remove potential spikes. We then co-added the spectra and
 improved the baseline solution by subtracting the linear
 interpolation of median values computed every 50 MHz over 100 MHz.


Although we are confident that there are no other species remaining to be detected 
in the survey data, it is not yet entirely free of artifacts,
for instance unexplained features, always downward-going, that are too broad to
be real absorption in the interstellar medium and are not always present in all 
tunings. As a check on the validity
of the data acquistion path we measured the integrated absorption for 
previously detected lines toward \bll\ and 3C111  and a comparison of new and 
old results is shown at top in Fig. 1.  Above an equivalent width of 0.3 MHz 
(1 \kms\ at 100 GHz) there 
is a slight tendency for the new results to be smaller, perhaps as the result of
under-resolving the stronger lines, hence underestimating their opacity.  
The SO 3-2 line at 99.3 GHz called out in Fig. 1 
falls nearly on the join of two tunings where the system temperature 
may not have been reliably inferred by the data reduction software and 
was the only feature so affected.

\subsection{Survey properties: rms}

Figure 1 at bottom shows the rms line/continuum noise for \bll\ and 3C111 across the 
frequency band surveyed.  To make these plots, we calculated a running rms over 8-channel 
intervals in line spectra divided by the continuum and
averaged this over absorption-free regions of (approximately) fixed mean rms, determined
by visual inspection.  When the running mean rms changed beyond twice the expected variance 
a new data point was created.
The rms increases at the extreme of the low end because less time was spent there and 
more broadly at the high end because of the increased atmospheric opacity and consequent 
increase of system temperature.  Inside
the extremes there are patches in the survey with noticeably higher rms. These are caused
by some combination of smaller integration time/higher system temperature and artifacts in
 the IF.
Some of these artifacts are narrow and resemble astronomical spectral lines while others are
much broader.  By careful comparison of tunings and polarizations we concluded that none
of the problematic features result from true astronomical signals.

Toward \bll\ $0.0015  <\sigma < 0.002$ over the region 85 - 110 
GHz leading to a single-channel detection sensitivity 3$\sigma$/\EBV$ \le$ 0.02/mag
 and a $3-\sigma$ detection sensitivity $\le 0.023 \sqrt(dV)$/mag when integrated
over a velocity interval dV (\kms) at 100 GHz.  The same  quantities obtained
toward 3C111 were $0.004 < \sigma < 0.006$ and 3$\sigma$/\EBV$ \le$ 0.011/mag for
the single-channel sensitivities and a $3-\sigma$ velocity-integrated sensitivity 
$\le 0.014 \sqrt(dV)/$mag at 100 GHz.

Thus, although the intrinsic channel-channel rms noise is considerably smaller toward 
\bll\ owing to its higher flux during the period of observations, the signal/noise of
the new detections is better toward 3C111 owing to the higher column density.
Nonetheless, the increased sensitivity toward 3C111 did not result in the 
detection of any additional species beyond those also seen toward \bll, 
even tentatively.  Future observations that seek to improve significantly on 
the current results will have to have rms $<<$ 0.002 at \EBV\ 
= 0.32 mag and rms/\EBV\ $<<$ 0.003 mag $^{-1}$ more generally.   

\subsection{Improved results for SiO and other previously-detected species}

\bll\ was not included in our earlier survey of SiO J=2-1 absorption \citep{LucLis00}
so the spectrum in Fig. 2 represents the first detection of SiO in this direction;
the SiO spectrum of NRAO150 is superior to that in the earlier work.   SiO
is one of several species (eg N$_2$H\p) whose abundances, either detections or upper 
limits, warrant rediscussion on the basis of the sensitivity achieved in the current work.
This work is in progress.

\begin{figure*}
\includegraphics[height=10cm]{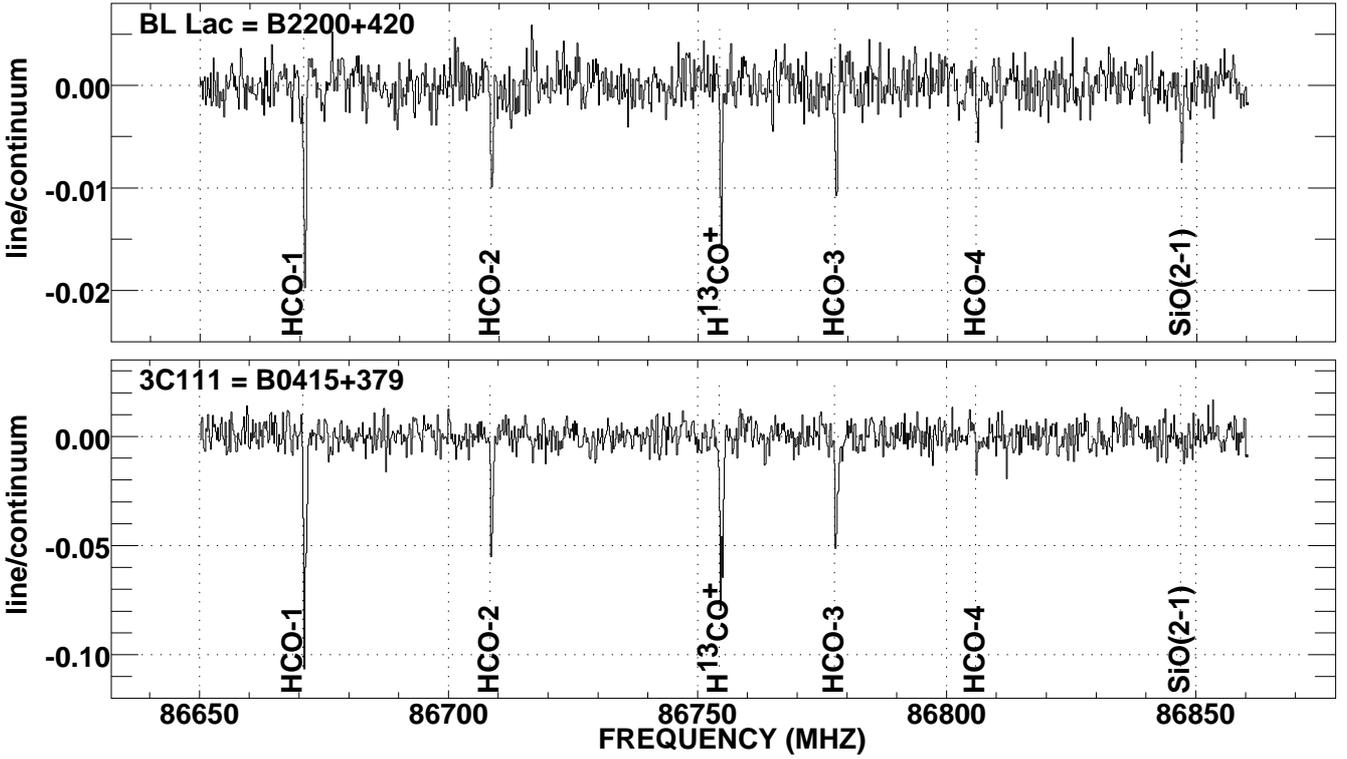}
  \caption[]{Spectra of \bll\ and 3C111 in the region of the HCO quartet.  
 The spectral resolution is 0.195 MHz.  The approximate positions of HCO, \hthcop\ and SiO
  J=2-1 transitions are marked.  For spectroscopic data, see Table 2.}
\end{figure*}

\begin{figure*}
\includegraphics[height=8cm]{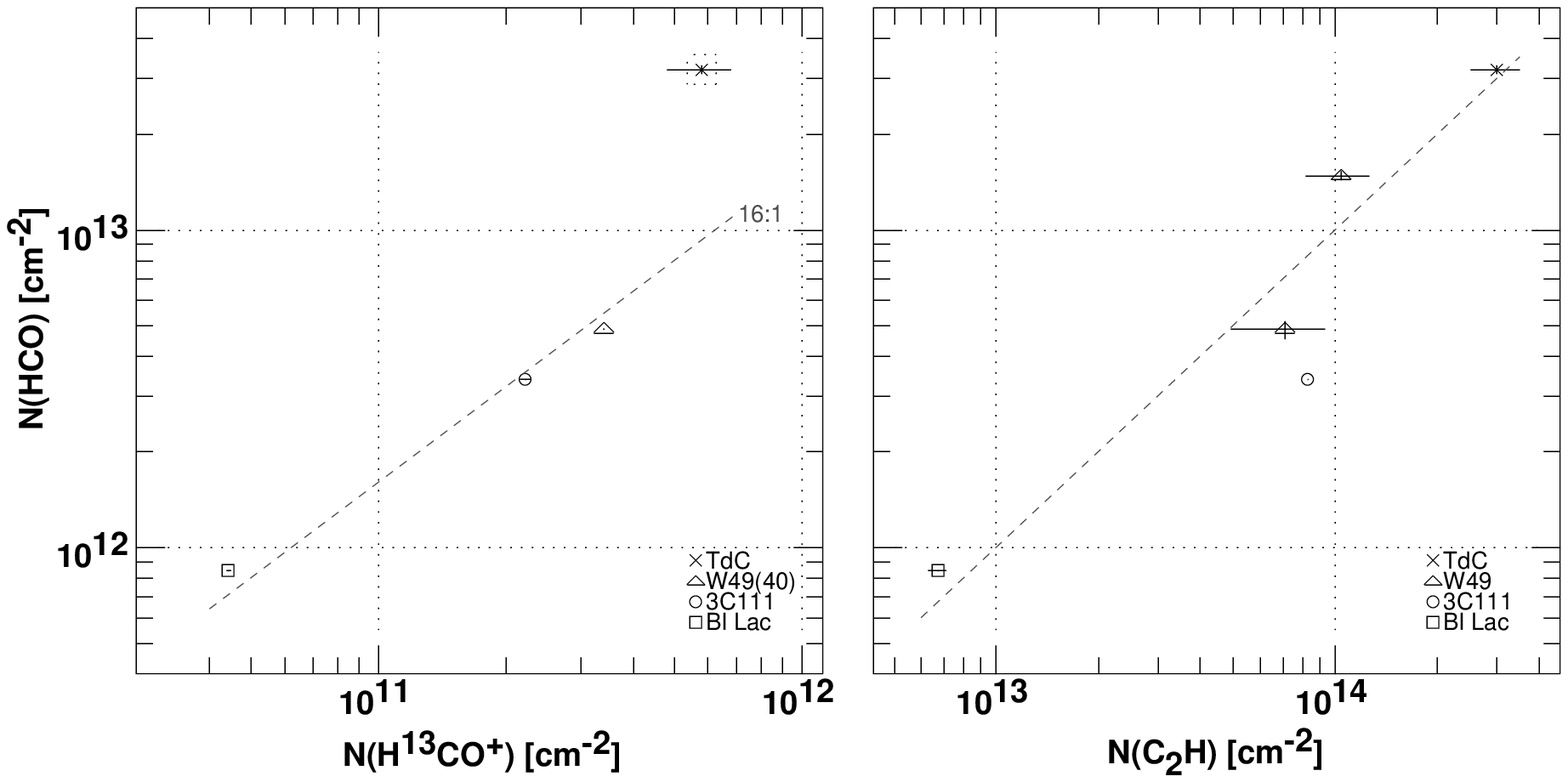}
  \caption[]{Column density of HCO vs. \hthcop\ (left) and \cch\ (right).  For W49 
 only the ``spiral arm'' feature
  at 40 \kms\ is shown at left using data from the present work; at right the \cch\
  data of \cite{GodFal+10} is used for the 40 \kms\ and 60 \kms\ features. Data 
 from the Horsehead PDR (TdC) are from \cite{GerGoi+09}. The dashed lines have
 power-law slope of unity, with proportions of 16:1 and 1:10 at left and right,
 respectively.  Error bars in this and all other plots are $1\sigma$ statistical
 uncertainties.}
\end{figure*}
 
\begin{table*}
\caption[]{Sightlines and continuum targets}
{
\small
\begin{tabular}{lccccccccccc}
\hline
Target & l  &  b & \EBV$^a$ & Flux & rms$^b$ & v & N(HCO) &N({\it c}-C$_3$H)& N(CF$^+$)& N(H$^{13}$CO\p) & N(\cch) \\ 
&  \degr  & \degr  & mag  & Jy &10$^{-3}$  &\kms & $10^{12}\pcc$ & $10^{11}\pcc$ & $10^{11}\pcc$ & $10^{11}\pcc$& $10^{13}\pcc$ \\
\hline
B0415+379 & 161.67 & -8.82 & 1.65 & 3$^c$ & 4-6 & -1 & 3.38(0.01) & 4.63(0.18)& 7.9(1.6)& 2.22& 8.29\\ 
B2200+420 & 92.59 & -10.44 & 0.32 & 15$^c$& 1.5-2 & -1 &0.85(0.02) &  1.62(0.05) &4.0(0.7)&0.42 & 0.67 \\
W49 & 43.17 & 0.01 & & 2.5 & 7 & 40   & 4.86(0.32) &&&3.40 &7.10\\ 
W49 & 43.17 & 0.01 & & 2.5 & 7 & 51-66 & 14.8(0.5) &&&&10.4\\ 
\hline
\end{tabular}}
\\
$^a$from \cite{SchFin+98} \\
$^b$ see Section 2.2 concerning the channel-channel line/continuum rms in the spectral sweep \\
$^c$ approximate, derived from pointing scans assuming 6 Jy Kelvin$^{-1}$ \\
\end{table*}

\begin{figure*}
\includegraphics[height=11.6cm]{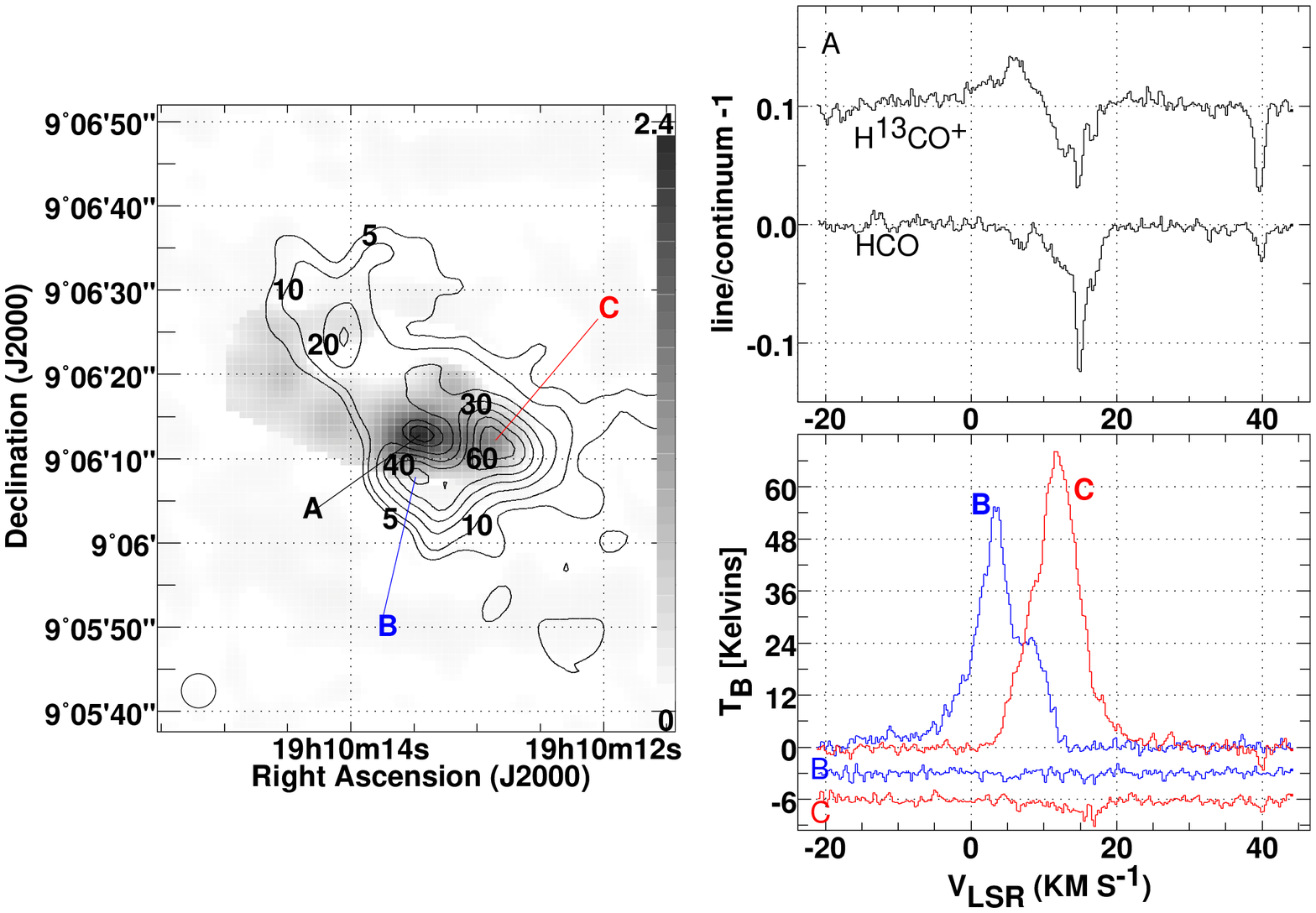}
  \caption[]{W49 continuum, HCO F=1-1 and \hthcop\ emission and absorption spectra.  Left: 
  86.8 GHz continuum (grayscale; peak = 2.5 Jy) and \hthcop\ emission (contours of 
  integrated brightness  temperature in units of Kelvins \kms).  The 4\arcsec\
 synthesized HPBW is shown inset at lower left.  
  Right top: spectra of \hthcop\ and HCO toward
  the continuum peak labelled position ``A'' at left. Right bottom: emission spectra
 at off-continuum positions ``B'' (blue in electronic media) and ``C'' (red). 
  Spectra showing strong emission are \hthcop, those without are spectra of HCO.}
\end{figure*}

\subsection{Aperture synthesis of W49 in HCO and \hthcop\ at the PdBI}

We observed the 86.774 GHz HCO F=1-1 line at the Plateau de Bure Interferometer for 
7 hours in September 2007 in the 
D configuration (baselines 25 - 100 m) and in 2008 April for 9 hours in the C
configuration (25 - 175 m) to achieve a combined spatial resolution of 
4.0\arcsec\ over a single primary beam pointing.  The spectral resolution was 
270 kHz or 0.93 \kms\ and the rms noise in the line/continuum ratio at 0 
absorption was 0.007 with 4.2 hours of usable integration time.  
The LO was centered slightly too low in velocity for simultaneous coverage of \hthcop\
in all of the ``spiral arm'' features that are seen in absorption in this direction
but the strong absorption at 40 \kms\ was captured.  All the previously known
spiral arm features were seen in HCO.

\section{HCO and its detection in diffuse clouds}

HCO was studied extensively at H II region-molecular cloud interfaces by Snyder and 
Schenewerk and their collaborators \citep{SchJew+88}, including Ori B and W49
but its presence in diffuse and translucent clouds has not previously been noticed.  
Table 4 of \cite{SchJew+88} has N(HCO)/N(\h13cop) = 1-3 generally and 10 toward 
NGC2024 but they did not derive column densities for either species individually.

The spectra of HCO toward \bll\ and 3C111 
are shown in Fig. 2 and toward W49 in Figure 4.   The transitions 
observed and their transition probabilities are given in Table 2;
the observed line depths are as expected in LTE.  The HCO column densities
quoted here in Table 1 were derived from a simultaneous gaussian fit to all 
observed transitions assuming that the excitation is in equilibrium with the cosmic
microwave background.  We find N(HCO)/N(H$^{13}$CO\p) = 16 toward \bll, 3C111 and the 
40 km/s spiral arm cloud toward W49, much larger than the values 1-3 seen in the 
denser regions studied by Schenewerk et al (1988).  The abundance of HCO
relative to molecular hydrogen from our work is X(HCO) = $8\times10^{-10}$ if
N(H$^{13}$CO\p)/N(\hcop) = 1/60 and X(\hcop) =   $3\times10^{-9}$.  This is
about 10 times higher than in the dark clouds TMC-1 or B1, according to 
recent unpublished results (Marcelino and Bacmann 2013, private communication).

Figure 3 at right shows a comparison of N(HCO) with N(\cch) in the diffuse and
spiral arm clouds at 40 and 60 \kms\ toward W49 using the \cch\ column densities
of \cite{GodFal+10}, and toward the Horsehead.  Most of these have abundance ratios 
N(HCO)/N(\cch) $\approx$ 0.1 with a three times smaller value toward 3C111, which has
a higher than normal ratio N(\cch)/N(\hcop).  For the Horsehead the peaks of the 
spatial distributions of \cch\ and HCO coincide just inside the illuminated edge 
of the PDR while the peak of \hcop\ lies further inside the neutral region. 
The Horsehead has a much higher ratio of N(HCO)/N(\hthcop) in Fig. 3 at left 
because the edge-on viewing geometry allows us to distinguish the HCO and \hcop\ 
peaks. 
If the Horsehead were observed face-on its HCO/\hthcop\ ratio would most likely
resemble much more closely that seen in the diffuse and spiral arm clouds
because all of the material would be observed along the same sighlines.
Whether such a segregation occurs in the diffuse and spiral arm clouds is 
impossible to discern.

Coincidence of the HCO and hydrocarbon peaks in the Horsehead may be ascribed to 
the formation mechanism for HCO, O + C\HH\ $\rightarrow$ HCO + H: \cite{GerGoi+09}
found that a rapid rate for this reaction was required to explain the observed
abundances.  In turn, C\HH\ formation is initiated by the endothermic reaction 
C\p\ + \HH\ but C\p\ does not survive into dense molecular gas after carbon
is converted to CO.  Therefore HCO is most properly viewed as existing in interface 
regions where the density is comparatively high and a substantial \HH\ fraction is 
present but carbon remains largely in the form of C\p.

\begin{figure*}
\includegraphics[height=10cm]{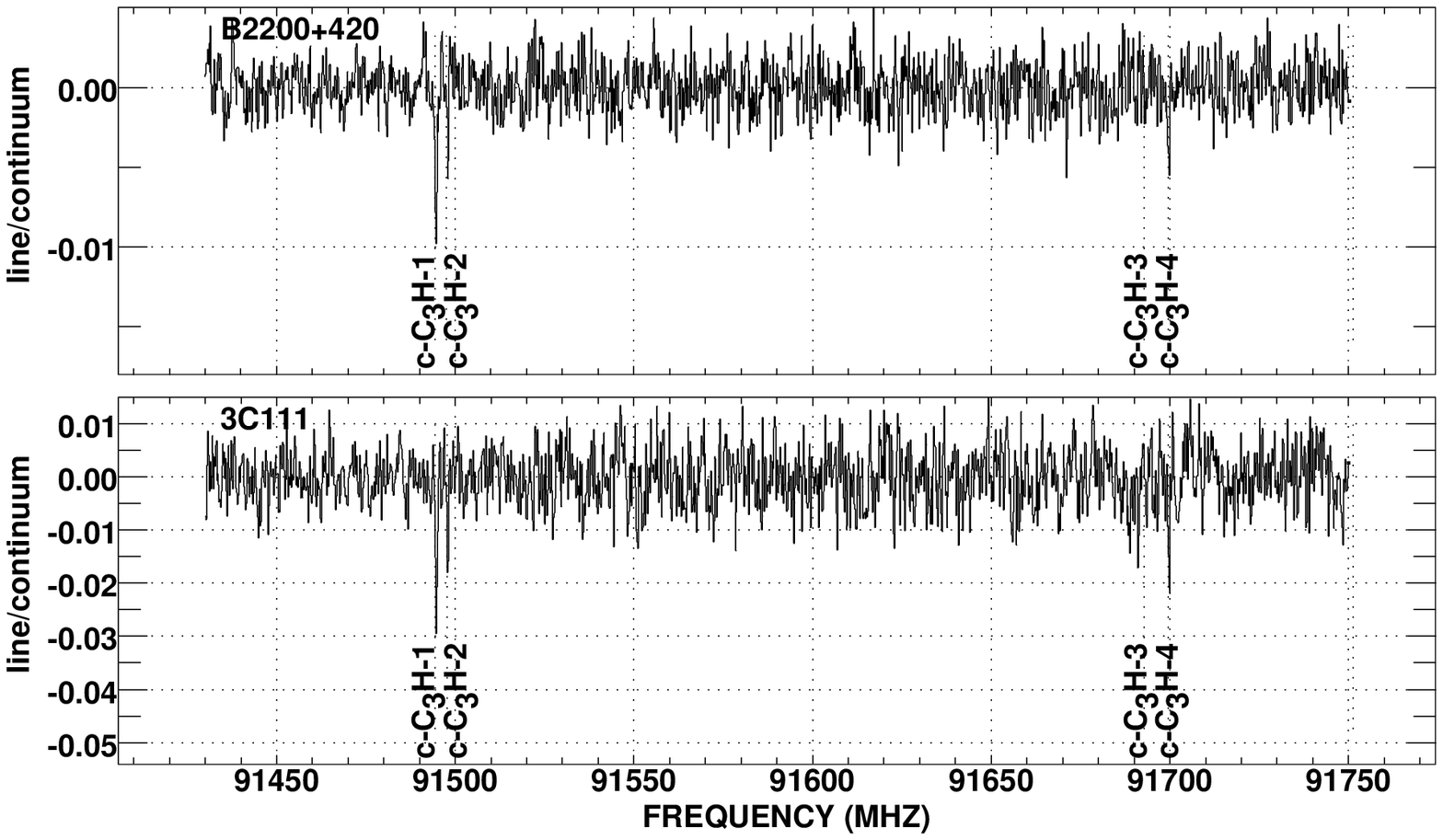}
  \caption[]{Spectra of \bll\ and 3C111 in the region of the \cc3h\ quartet.  
 The spectral resolution is 0.195 MHz.  For spectroscopic data, see Table 3.}
\end{figure*}

Figure 4 illustrates how HCO is present in the gas within the W49 HII region-molecular 
cloud complex as a whole but does not persist into the denser neutral regions.
The maps at left superpose contours of integrated \hthcop\ emission upon the 
continuum in greyscale, noting three positions A (the continuum), B and C 
at which line profiles are shown at right.  \hthcop\ at positions B and C 
(lower right) shows blue and red-shifted emission
features, respectively (with slight overlap), corresponding to the blue-shifted
\hthcop\ emission and red-shifted absorption seen toward the continuuum at position
A.  Thus the overall morphology can be explained by a disk geometry whose blue-shifted
southeastern part is situated behind the continuum, so representing infalling
material.  By contrast, HCO at position A
toward the continuum shows strong red-shifted absorption but no blue-shifted emission 
from the denser gas behind the continuum.
Likewise, HCO shows no detectable emission off the continuum at positions B or C.
HCO exists only in lower density portions of the molecular gas where the abundance of
C\p\ is relatively large but the molecular excitation of HCO is relatively weak.

\begin{table}
\caption[]{HCO 1$_{01}$-0$_{00}$ transitions observed and column density-optical 
depth conversions$^a$ }
{
\small
\begin{tabular}{lcccc}
\hline
Frequency&F$^\prime$ - F & J$^\prime$ - J & \AF  & N(X)/$\int\tau dv^b $ \\
 MHz      &        &  & 10$^{-6}$\ps  &   $\pcc (\kms)^{-1}$ \\
\hline
86670.76 & 2-1 & 3/2-1/2 & 4.69& $2.26\times10^{13} $ \\
86708.36 & 1-0 & 3/2-1/2 & 4.60& $3.84\times10^{13} $ \\
86774.46$^c$ & 1-1 & 1/2-1/2 & 4.61& $3.84\times10^{13} $ \\
86805.75 & 0-1 & 1/2-1/2 & 4.72& $1.13\times10^{14} $ \\
\hline
\end{tabular}}
\\
$^a$ Spectroscopic data from {\tt www.splatalogue.net} see \cite{BlaSas+84}\\
$^b$ Assuming excitation in equilibrium with the cosmic microwave background \\
$^c$ This component was observed toward W49 \\

\end{table}

\begin{table}
\caption[]{\cc3h\  2$_{12}$-1$_{11}$ transitions observed and column density-optical 
depth conversions$^a$ }
{
\small
\begin{tabular}{lcccc}
\hline
Frequency &F$^\prime$ - F & J$^\prime$ - J & \AF & N(X)/$\int\tau dv^b $ \\
 MHz      &        &  & 10$^{-5}$\ps  &   $\pcc (\kms)^{-1}$ \\
\hline
91494.35 & 3-2 & 5/2-3/2 & 1.545& $1.02\times10^{13} $ \\
91497.61 & 2-1 & 5/2-3/2 & 1.545 & $1.42\times10^{13} $ \\
91692.75 & 1-0 & 3/2-1/2 & 1.545& $2.39\times10^{13} $ \\
91699.47 & 2-1 & 3/2-1/2 & 1.545& $1.43\times10^{13} $ \\
\hline
\end{tabular}}
\\
$^a$ Spectroscopic data from \cite{ChaShi+07} \\ 
$^b$ Assuming excitation in equilibrium with the cosmic microwave background \\
\end{table}

\section{\cc3h\ and its detection in diffuse clouds}

The discovery spectra of \cc3h\ in diffuse and translucent clouds are shown in 
Fig. 5. 
The transitions observed and their transition probabilities are given in Table 3;
the observed line depths are as expected in LTE.  The column densities
quoted here in Table 1 were derived from a simultaneous gaussian fit to all 
observed transitions assuming that all excitation is in equilibrium with the cosmic
microwave background. 

Fig. 6 compares the observed column densities of \cc3h\ with those of \hthcop, 
{\it ortho-c-}\c3h2\
and \cch.  The values quoted for the Horsehead in Fig. 6  are from the singledish IRAM 
30m observations of \cite{PetGra+12} at lower spatial resolution than those shown in Fig. 4.
The quasar absorption sightlines and TMC-1 have similar N(\cc3h)/N(\hthcop) $\approx 3$, 
N(\cc3h)/N(o-c-\c3h2) $\approx 0.1$ and  N(\cc3h)/N(\hthcop) $\approx 0.01$. 
N(\cc3h) for the Horsehead is high by a factor two-four with regard to \hthcop\
and {\it c-}\c3h2, and perhaps somewhat high with respect to \cch\ as well.

\cite{ManWoo90} noted that the relative abundances of \cc3h\ and c-\c3h2\ varied little
with environment, whether in dark or giant molecular clouds, HII region-molecular
cloud complexes, etc.  Our data greatly extend this conclusion downward in 
column density and across chemical families.  At left in Fig. 6 it is seen that 
N(\cc3h)/N(\hthcop) = 3 toward TMC-1 and the two blazars observed here, over a range of 
a factor 40 in column density.  At right the regression lines have power-law slope 0.80,
so that N(\cc3h)/N(\c3h2) and N(\cc3h)/N(\cch) vary by a factor $\pm 2$ about a common mean
over a factor 100 in N({\it c-}\c3h2) and N(\cch).

\begin{figure*}
\includegraphics[height=8cm]{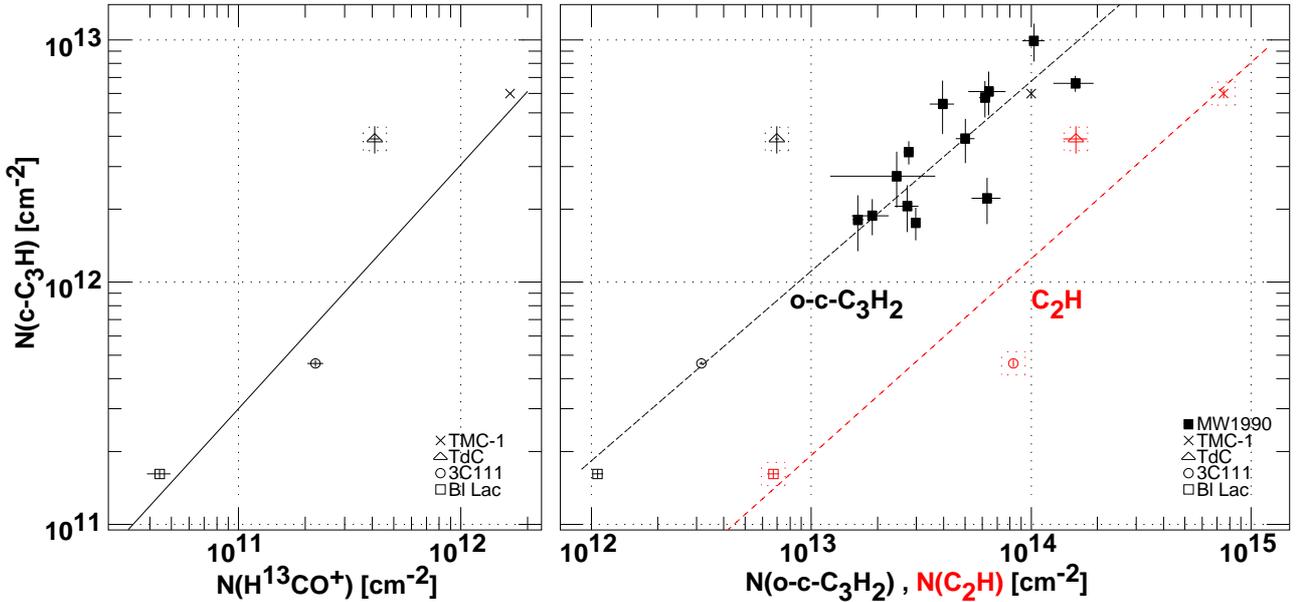}
  \caption[]{Column density of \cc3h\ vs. \hthcop\ (at left), and with \c3h2\ and \cch\ (right).
  Data for the hydrocarbons toward the Horsehead nebula (TdC) are from \cite{PetGra+12}.
  Data for TMC-1 are from \cite{OhiIrv+92}.  The \cc3h\ survey data of \cite{ManWoo90} 
  are shown at right, labelled ``MW1990''. The solid line at left has power-law slope 1;
  at right, both lines have slope 0.8.}
\end{figure*}

From the lack of variability in their abundance ratio, \cite{ManWoo90} inferred that 
\cc3h\ and \c3h2\ shared a common chemical  antecedent C$_3${H$_3$}$^+$.  Although this 
may be true, Fig. 6 suggests that \cc3h\ is simply another species, like \hcop, \cch, and 
\c3h2, whose abundance relative to \HH\ varies little in the ISM at large.

\section{CF\p\ and its detection in diffuse clouds}

CF\p\ has previously been observed only in regions of much higher
density, the Orion Bar \citep{NeuSch+06} and the Horsehead 
\citep{GuzPet+12,GuzRou+12}.
The discovery spectra of CF\p\ in diffuse clouds are shown in Fig. 7 and 
the transitions observed and their transition probabilities are given in Table 4.  
The low spectral resolution of the sweep data,
combined with the 0.34 MHz or 1.01 \kms\ separation of the two hyperfine
components prevents a reliable determination of the relative strengths of the 
kinematic components toward 3C111.  The column densities quoted here in 
Table 1 were derived by integrating over the observed profiles. 

Fig. 8 shows that N(CF\p) increases with N(\hthcop) and N(\cch), although
more slowly than linearly (power law slopes 0.6 and 0.4 respectively).  
Fig. 8 also shows the prediction of \cite{NeuWol09} that 
N(CF\p) $\approx 4.5 \times 10^{11}\pcc$ for \AV\ $\ga 1$
with most fluorine in HF in the \HH-bearing regions.  The  conversion of 
carbon to CO and the abrupt depletion of gas phase fluorine at \AV\ $>$ 1 mag 
both act to confine CF\p\ at \AV\ $<$ 1 mag and to limit N(CF\p) in the models 
of \cite{NeuWol09}.

CF\p\ forms from HF {\it via} C\p\ + HF $\rightarrow$ CF\p\ + H with a 
rate constant k$_1= 7.2\times10^{-9}(T/300)^{-0.15}{\rm cm}^3$\ps\ and 
is destroyed by recombination with electrons with a recombination coefficient 
$\alpha_1= 5.3\times10^{-8}(T/300)^{-0.8}{\rm cm}^3$\ps.  If C\p\ is
the dominant source of free electrons it follows that n(CF\p)/n(HF) = 
0.136 (T/300)$^{0.65}$ with most gas-phase fluorine in HF.

We know that C\p\ survives as the dominant form of carbon at and well beyond 
\AV\ = 1 mag because of the strong fractionation of \THC\ into \coth\ 
that is seen toward \bll\ and 3C111 \citep{LisLuc98}; substantial
amounts of gas-phase fluorine in the form of HF must also survive beyond 
\AV\ = 1 mag to provide the behaviour seen in Fig. 8.  If
the fractional abundances of \hcop\ and \cch\ with respect to \HH\
are about constant and the kinetic temperature does not vary
substantially, the depletion of gas phase 
fluorine (HF) would vary approximately as \AV$^{-1/2}$.  If
the temperature were assumed to decline at higher \AV, the depletion of fluorine
would have to be even weaker in order to compensate for the smaller 
equilibrium CF\p/HF ratio, which varies as T$^{0.65}$ according to the
simple chemical scheme detailed above \citep{NeuWol09}.

\section{Summary}

We surveyed the 84 - 116 GHz 3mm spectrum in absorption against the compact 
extragalactic  mm-continuum  sources \bll\ and 3C111 as noted in Table 1 at 
195 kHz spectral resolution (0.6 \kms\ at 100 GHz), achieving an rms noise 
level $\delta_\tau \approx 0.002$ at \EBV\ = 0.32 mag (\AV\ = 1 mag) toward
\bll\ and  $\delta_\tau$/\EBV\ $\approx\ 0.003$ mag$^{-1}$ overall.  HCO, \cc3h\
and CF\p\ were detected in absorption toward both sources and HCO was also found 
in the diffuse ``spiral-arm'' clouds in the galactic plane in front of W49.
We discussed observational aspects of their chemistry in Sections 3-5.

\cc3h\ is notable for having a nearly fixed abundance with respect to \hcop,
\cch\ and c-\c3h2\ over a wide range of column density, even across the divide
between the diffuse and dark or giant molecular clouds.  The increase in
N(CF\p) beyond \AV\ = 1 mag shows that both fluorine and C\p\ are abundant in the 
gas phase at such extinctions but conclusions about the depletion of
fluorine are sensitive to assumptions about the temperature profile.  
The relative abundance ratio  N(HCO)/N(\HH) is
higher in diffuse than dark or dense molecular gas, consistent with its prior
interpretation as a species requiring both C\p\ and \HH\ for its formation via
the reaction  O + C\HH\ $\rightarrow$ HCO + H.

No other new species are present in the spectra toward \bll\ (\AV = 1 mag) at a 5-sigma 
optical depth limit of 0.01 in any single feature, suggesting that the hunt for new 
species in diffuse clouds in the radio domain will be most profitably conducted in 
the cm-wave region below 30 GHz where the more heavily-populated lower-lying 
energy levels of heavier polyatomic molecules are most readily accessible.

\begin{figure}
\includegraphics[height=7cm]{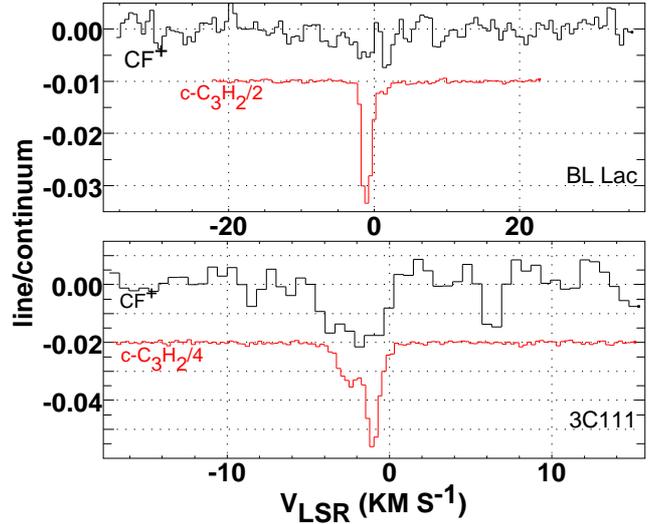}
  \caption[]{CF\p\ spectra toward \bll\ (upper) and 3C111 (lower).  The velocity
axis corresponds to the mean LTE intensity-weighted centroid of the two hyperfine
 components noted in Table 4. Shown 
 for comparison are scaled spectra of the 18.3 GHz transition of c-\c3h2\ 
  from \cite{LisSon+12}. For spectroscopic data, see Table 4.}
\end{figure}

\begin{figure}
\includegraphics[height=6.6cm]{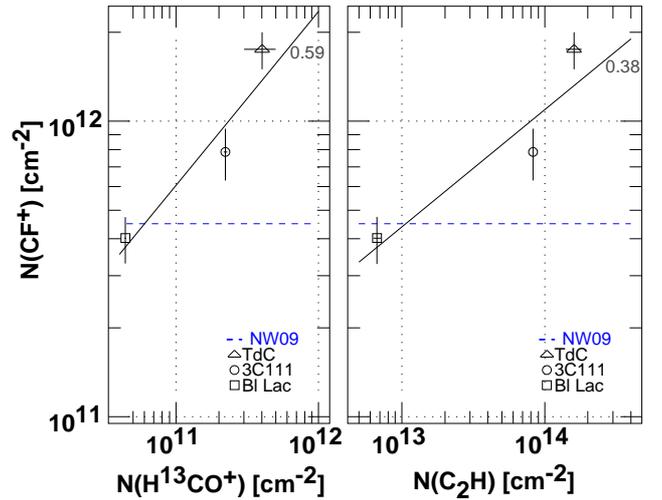}
  \caption[]{Column density of CF\p\ vs. \hthcop\ (at left) and  \cch\ (right).
  Data for the Horsehead nebula (TdC) are from \cite{GuzPet+12}.  The prediction
 N(CF\p) $\approx 4.5 \times 10^{11}\pcc$ at \AV\ $\ga$ 1 mag of \cite{NeuWol09}
 is shown as a blue dashed horizontal line.  The power law slopes of the best-fit
 lines (0.59, 0.38) are noted.}
\end{figure}

\begin{table}
\caption[]{CF\p\ transitions observed and column density-optical 
depth conversions$^a$}
{
\small
\begin{tabular}{lcccc}
\hline
Frequency &F$^\prime$ - F & J$^\prime$ - J & \AF     & N(X)/$\int\tau dv^b $ \\
 MHz      &        &  & 10$^{-6}$\ps  &   $\pcc (\kms)^{-1}$ \\
\hline
102587.189 & 1/2-1/2 & 1-0 & 4.82 & $4.04\times10^{13} $ \\
102587.533 & 3/2-1/2 & 1-0 & 4.82 & $2.02\times10^{13} $ \\
\hline
\end{tabular}}
\\
$^a$ Spectroscopic data from \cite{GuzRou+12} and {\tt www.splatalogue.net}\\
$^b$ Assuming excitation in equilibrium with the cosmic microwave background \\
\end{table}

\begin{appendix}

\section{Upper limits on undetected species}

In most cases we have examined, the upper limits we can derive for undetected species 
are not sufficiently sensitive to be interesting.  For instance, we derive 
$3\sigma$ upper limits 
N(\lc3h\p) $< 2.3 \times 10^{11}\pcc$ and N(\lc3h\p) $ < 6.0 \times 10^{11} \pcc$  
toward \bll\ and 3C111, respectively, but these are at most barely below the value  
N(\lc3h\p) = $4.8 \times 10^{11} \pcc$ for the Horsehead \citep{PetGra+12}
whose column densities are generally 3-10 times larger.  For the neutral \lc3h\
the upper limits that may be derived from this work are not competitive with those
accessible at lower frequency using the VLA and likewise with \cfh\ \citep{LisSon+12}.

One species that is better-constrained is \HH CS\ whose J=3-2 transitions appear
at 101.5 and 104.6 GHz.  We find $3 \sigma$ limits N(\hhco)/N(\HH CS) $> 11 \pm 2.5$ and 
N(\hhco)/N(\HH CS) $> 32 \pm 8$ toward \bll\ and 3C111 respectively using the
prior results of \cite{LisLuc+06}, compared to N(\hhco)/N(\HH CS) = 7 toward TMC-1
\citep{OhiIrv+92}.

\end{appendix}

\begin{acknowledgements}
The National Radio Astronomy Observatory is a facility of the National Science Foundation 
operated under cooperative agreement by Associated Universities, Inc.
IRAM is operated by CNRS (France), the MPG (Germany) and the IGN (Spain). 
This work was partly funded by grant  ANR-09-BLAN-0231-01 from the French 
{\it Agence Nationale de la Recherche} as part of the SCHISM project.
We thank the editor, Malcolm Walmseley, and the anonymous referee for 
their comments.

\end{acknowledgements}
 
\bibliographystyle{apj}


\end{document}